%Paper: hep-th/9507103
%From: H.Suzuki <hsuzuki@mito.ipc.ibaraki.ac.jp>
%Date: Thu, 20 Jul 95 18:18:08 JST

\input phyzzx
%%%%%%%%%%%%%%%%%%%%%%%%%%%%%%%%%%%%%%%%%%%%%%%%%%%%%%%%%%%%%%%%%%%%%%%%%%%%%%
\REF\BRE{
E. Br\'ezin, G. Parisi and J. Zinn-Justin, {\sl Phys. Rev.\/} {\bf D16}
(1977) 408.}
\REF\AOY{
H. Aoyama, {\sl Mod. Phys. Lett.\/} {\bf A7} (1992) 1337;\hfill\break
H. Aoyama and A. M. Tamra, {\sl Nucl. Phys.\/} {\bf B384} (1992) 229.}
\REF\MOR{
W. Moretti, Universit\`a di Genova thesis (unpublished); A. Bini,
Universit\`a di Genova thesis (unpublished).}
\REF\ZIN{
J. Zinn-Justin, {\sl Nucl. Phys.\/} {\bf B192} (1981) 125;
{\sl Nucl. Phys.\/} {\bf B218} (1983) 333.}
\REF\BUC{
I. R. C. Buckley, A. Duncan and H. F. Jones, {\sl Phys. Rev.\/} {\bf D47}
(1993) 2554;\hfill\break
A. Duncan and H. F. Jones, {\sl Phys. Rev.\/} {\bf D47} (1993) 2560.}
%
%%%%%%%%%%%%%%%%%%%%%%%%%%%%%%%%%%%%%%%%%%%%%%%%%%%%%%%%%%%%%%%%%%%%%%%%%
%
\pubnum={IU-MSTP/4; hep-th/9507103}
\date={July 1995}
\titlepage
\title{Calculation Rule for Aoyama--Tamra's Prescription
for Path Integral with Quantum Tunnelling}
\author{
Hiroshi Suzuki\foot{e-mail: hsuzuki@mito.ipc.ibaraki.ac.jp}}
\address{
Department of Physics, Ibaraki University, Mito 310, Japan}
\abstract{We derive a simple calculation rule for Aoyama--Tamra's
prescription for path integral with degenerated potential minima.
Non-perturbative corrections due to the restricted functional space
(fundamental region) can systematically be computed with this rule.
It becomes manifest that the prescription might give Borel summable
series for finite temperature (or volume) system with quantum tunneling,
while the advantage is lost at zero temperature (or infinite volume)
limit.}
\endpage
%%%%%%%%%%%%%%%%%%%%%%%%%%%%%%%%%%%%%%%%%%%%%%%%%%%%%%%%%%%%%%%%%%%%%%%%%%

\overfullrule=0pt
In quantum mechanics a perturbative expansion of the partition function
(or of Green functions, of energy levels) around a degenerated
potential minimum is known to be non-Borel summable, due to quantum
tunneling [\BRE]. This is also true for scalar field theories in finite
volume and non-Abelian gauge field theories. For such a system,
therefore it is a fundamental problem how to relate the true value
and the information of perturbation series.

Concerning this problem, Aoyama, and Aoyama and Tamra [\AOY] proposed
a prescription for path integral with a degenerated potential. Their
observation is following: Take an integral
$$
   z(g)={1\over2\sqrt{2\pi}}\int_{-\infty}^\infty d\phi\,
   \exp\left[-{1\over2}\phi^2(1-g\phi)^2\right].
\eqn\one
$$
We see $\lim_{g\to0}z(g)=1$ since the integrand has two Gaussian
peaks located at $\phi=0$ and $\phi=1/g$ in the weak coupling limit.
The standard perturbation series around $\phi=0$, on the other hand, gives
$$
   z(g)_{\rm pert.}={1\over2}+3g^2+105g^4+\cdots.
\eqn\two
$$
We realize the series is not even an asymptotic expansion in the sense that
$\lim_{g\to0}|z(g)-z(g)_{\rm pert.}|=1/2$ (where a termination of
the series is undestood). Also $\lim_{g\to0}z(g)=2z(0)$, thus
$z(g)$ has a discontinuity at $g=0$. The authors of [\AOY] identified this
discontinuity at $g=0$ as the origin of the non-Borel summability of
series expansion. To remedy this point they proposed to define the
integral instead by
$$
   Z(g)={1\over\sqrt{2\pi}}\int_{-\infty}^{1/2g} d\phi\,
   \exp\left[-{1\over2}\phi^2(1-g\phi)^2\right].
\eqn\three
$$
For $g$ real positive, $Z(g)=z(g)$, but $Z(0)=2z(0)$ thus $Z(g)$
(as a function of real non-negative $g$) is continuous at $g=0$
in contrast to $z(g)$. Though $Z(g)$ differs from the original $z(g)$
at $g=0$, if \three\ gives Borel summable series it is a great progress
as we are interested in $z(g)$ with $g$ real positive.

The prescription [\AOY] can be summarized as follows: Divide the
integration region (or functional space) appropriately according to the
symmetry of the potential ($\phi\leftrightarrow1/g-\phi$ in the above
example). Then multiply the number of the degeneracy
($2$ in the above example). The prescription can readily be generalized
to quantum mechanics and field theories (see below).

However at first glance it is not obvious how to systematically
approximate the prescription \three, or the quantum
mechanical analogue, since the integration range is restricted by
a non-trivial way. The building block of the perturbative
expansion\foot{By the perturbative expansion we mean an expansion
with respect to $g$ in the exponential of the integrand in \three,
while $g$ in the limit of the integration region is kept fixed.
The expansion {\sl coefficients\/} of the series therefore
depend on $g$.} is not the
standard Gaussian integral. Without an explicit method for computation,
calling a numerical evaluation, the proposal [\AOY] might not practical.

If we are interested only in the above integral, of course it is a
simple matter to evaluate it perturbatively. Following the standard procedure
and
completing the square, we have
$$
   Z(g)=2Z_0
   \exp\left(
   g{\partial^3\over\partial j^3}-{1\over2}g^2{\partial^4\over\partial j^4}
   \right)
   \exp\left({1\over2}j\Delta j\right)C(\bar j)\biggr|_{j=0},
\eqn\four
$$
where $Z_0=1/2$, $\Delta=1$, $\bar j=j$ and we have defined
$$
   C(j)=
   1
   -{1\over\sqrt{\pi}}\int_{\sqrt{c\over2}(1/(2g)-j)}^\infty dt\,e^{-t^2},
\eqn\five
$$
with $c=1$. We note
$$
   C(0)-1=-{1\over\sqrt{\pi}}{\rm Erfc}\left(\sqrt{c\over2}{1\over2g}\right),
\eqn\six
$$
and the $n$th derivative,
$$
   C^{(n)}(0)=\sqrt{c\over2\pi}e^{-c/(8g^2)}
   \sum_{r=0}^{[n/2-1/2]}{(-1)^{r+1}(n-1)!\over r!(n-2r-1)!}
   \left({c\over2}\right)^{n-r-1}\left({1\over g}\right)^{n-2r-1}.
\eqn\seven
$$
We see $C^{(n)}(0)$ is proportional to $\exp[-c/(8g^2)]$ reflecting its
non-perturbative nature. This is also true for $C(0)-1$. The formula
\five\ may be used for a systematic analytic evaluation and the
lower order terms read
$$
\eqalign{
   &Z(g)=1-{1\over\sqrt{\pi}}{\rm Erfc}\left({1\over\sqrt{2}2g}\right)
   +g\left(-2-{1\over4g^2}\right){1\over\sqrt{2\pi}}e^{-1/(8g^2)}
\cr
   &+g^2\left[
   6-{6\over\sqrt{\pi}}{\rm Erfc}\left({1\over\sqrt{2}2g}\right)
   +\left(-{21\over4g}+{1\over8g^3}-{1\over64g^5}\right)
   {1\over\sqrt{2\pi}}e^{-1/(8g^2)}
   \right]+\cdots.
\cr
}
\eqn\eight
$$
Note that terms odd in $g$ survive in this expansion. The expansion is
asymptotically equal to $1+6g^2+\cdots$ at $g\sim0$, as is expected, and
differs from the standard perturbative series \two\ (normalized by hand [\AOY])
only by non-perturbative amounts. The non-perturbative correction is expected
to
improve the nature of the perturbative series and in fact, the large order
behavior
of the expansion ($Z(g)=\sum_{n=0}^\infty k_n(g)g^n$) is estimated in [\AOY]:
$$
   k_n(g)\sim {(-1)^n\over\sqrt{2\pi}}4^ne^{-n/2}n^{(n-1)/2},
\eqn\nine
$$
for $n\gg1/g^2$, thus is Borel summable.

The prescription can naturally be extended to higher dimensional cases.
For quantum mechanics with a double well potential,
$V(\phi)=\phi^2(1-g\phi)^2/2$,
the prescription gives\foot{We impose a periodic boundary condition
on the period $[0,\beta]$}
$$
   Z(g)=2N\int_{\bar\phi<1/(2g)}{\cal D}\phi\,e^{-S[\phi,g]},
\eqn\ten
$$
where
$$
   S[\phi,g]=\int_0^\beta d\tau\,\left[{1\over2}\dot\phi^2
                                       +{1\over2}\phi^2(1-g\phi)^2\right],
\eqn\eleven
$$
and the average of $\phi$ is defined by\foot{It is possible [\MOR] to
generalize the constraint as $\bar\phi=\int d\tau\,\phi w/\beta$,
introducing a weight function $w(\tau)$, $\int d\tau\,w/\beta=1$.
Our result is accordingly modified as
$\bar j=\int d\tau\int d\tau'\,w\Delta j/\beta$ and $c=\beta/a$,
where $a=\int d\tau\int d\tau'\,w\Delta w/\beta$. Since
$a<\int d\tau\, w^2/\beta$, we have the same conclusion on
$\beta\to\infty$ limit when the squared integral of the
weight function is bounded by $\beta$.}
$$
   \bar\phi={1\over\beta}\int_0^\beta d\tau\,\phi(\tau).
\eqn\twelve
$$

In what follows we will show that the partition function $Z(g)$ defined
by \ten\ is given exactly the same formula as \four\ with the following
substitutions:
$$
\eqalign{
   &Z_0\to
   Z_0\equiv N\int{\cal D}\phi\,\exp\left[-\int_0^\beta
   \left({1\over2}\dot\phi^2+{1\over2}\phi^2\right)
   \right]
   ={e^{-\beta/2}\over1-e^{-\beta}},
\cr
   &\Delta\to\Delta(\tau-\tau')=
   {\cosh(\beta/2-|\tau-\tau'|)\over2\sinh(\beta/2)},
\cr
   &c\to\beta,
\cr
   &\bar j\to\bar j\equiv{1\over\beta}\int_0^\beta d\tau\,j(\tau).
\cr
}
\eqn\thirteen
$$
(Some abbreviations in \four\ are understood; the derivative with respect
to $j$ is replaced by a functional derivative with $j(\tau)$, the
product $j\Delta j$ implies double integral over $\tau$.)
Therefore we have a systematic computational rule also for the quantum
mechanical case.

Since
$$
   Z(g)=2N\exp\left[-\int_0^\beta
   \left(-g{\delta^3\over\delta j^3}+{1\over2}g^2{\delta^4\over\delta
j^4}\right)
   \right]
   \int_{\bar\phi<1/(2g)}{\cal D}\phi\,
   e^{-S[\phi,0]+\int_0^\beta d\tau\,j\phi}
   \biggr|_{j=0},
\eqn\fourteen
$$
it is sufficient to consider
$$
\eqalign{
   Z_0[j]&\equiv
   N\int_{\bar\phi<1/(2g)}{\cal D}\phi\,
   e^{-S[\phi,0]+\int_0^\beta d\tau\,j\phi}
\cr
   &=N\int{\cal D}\phi\,\theta\left({1\over2g}-\bar\phi\right)
   e^{-S[\phi,0]+\int_0^\beta d\tau\,j\phi}.
\cr
}
\eqn\fifteen
$$
We first note
$$
   \theta\left({1\over2g}-\bar\phi\right)=
   \int_{-\infty}^\infty{dk\over2\pi i}{1\over k-i\varepsilon}
   \exp\left[ik\left({1\over2g}-\bar\phi\right)\right],
\eqn\sixteen
$$
therefore
$$
\eqalign{
   &Z_0[j]
\cr
   &=Z_0
   \int_{-\infty}^\infty{dk\over2\pi i}{e^{ik/(2g)}\over k-i\varepsilon}
   \exp\left\{
   {1\over2}\int_0^\beta d\tau\int_0^\beta d\tau'
   \left[j(\tau)-i{k\over\beta}\right]\Delta(\tau-\tau')
   \left[j(\tau')-i{k\over\beta}\right]
   \right\}.
\cr
}
\eqn\seventeen
$$
By using
$$
   \int_0^\beta d\tau'\Delta(\tau-\tau')=1,
\eqn\eighteen
$$
we obtain
$$
\eqalign{
   &Z_0[j]
\cr
   &=Z_0
   \exp\left[{1\over2}
   \int_0^\beta d\tau\int_0^\beta
d\tau'j(\tau)\Delta(\tau-\tau')j(\tau')\right]
   \int_{-\infty}^\infty{dk\over2\pi i}{e^{ik/(2g)}\over k-i\varepsilon}
   \exp\left(-ik\bar j-{k^2\over2\beta}\right).
\cr}
\eqn\nineteen
$$
Finally we use
$$
   \int_{-\infty}^\infty{dk\over2\pi i}{1\over k-i\varepsilon}
   \cos k\left({1\over2g}-\bar j\right)
   e^{-k^2/(2\beta)}={1\over2},
\eqn\twenty
$$
and
$$
   \int_{-\infty}^\infty{dk\over2\pi i}{1\over k-i\varepsilon}
   i\sin k\left({1\over2g}-\bar j\right)
   e^{-k^2/(2\beta)}=
   {1\over2}-{1\over\sqrt{\pi}}\int_{\sqrt{\beta\over2}
   (1/(2g)-\bar j)}^\infty dt\,e^{-t^2}.
\eqn\twentyone
$$
Convining \nineteen, \twenty\ and \twentyone\ and \fourteen,
we see the above statement: the partition function $Z(g)$ is given
by \four\ with the simple substitution \thirteen.

Clearly the same formula with trivial modifications also holds for a
(Eucledian) scalar field theory with the double well potential in finite volume
(especially $\beta$ is replaced by the volume). As is discussed in
[\AOY], the prescription gives Borel summable series even for quantum
mechanical models and the above rule might be used as a systematic
computation method.

However from the above demonstration, it becomes
obvious the advantage of the prescription is lost for $\beta\to\infty$
(or infinite volume limit)\foot{This fact is already briefly mentioned
in [\AOY] and also expected from another view point: The constraint
for the fundamental region $\bar\phi<1/(2g)$ cannot ``kill'' the
instanton--anti-instanton (${\rm I}$-$\bar{\rm I}$) configulation
when $\beta\to\infty$. The perturbative series around
${\rm I}$-$\bar{\rm I}$ is known to be non-Borel summable and the
imaginary part of the Borel sum should be cancelled by the one of
the perturbation series around trivial vacuum [\ZIN]. The prescription
therefore cannot give a Borel summable series.
The author thanks K. Konishi for explaining this point.}.
Namely in the formulas \six\ and \seven, we have $c\to\infty$ in such a
limit. All the non-perturbative corrections therefore vanish and the series
exactly reduces to the standard perturbation series, for which we know
the non-Borel summability.

In conclusion the prescription proposed by [\AOY] can systematically be
treated by an analytical method and may be used as an improved
perturbation scheme for finite volume system (considerations on the
renormalizability and the unitarity might be interesting).
The advantage however is lost for infinite volume limit.
The situation is quite similar to the delta expansion [\BUC]
directly applied to the partition function.

The author thanks K. Konishi for a careful reading of the manuscript and
imforming similar results are found in [\MOR].

\refout
%%%%%%%%%%%%%%%%%%%%%%%%%%%%%%%%%%%%%%%%%%%%%%%%%%%%%%%%%%%%%%%%%%%%%%%%
\bye